\documentclass[12pt]{article}
\usepackage{graphicx}
\usepackage{lscape}
\usepackage{citesort}
\usepackage{amssymb}
\usepackage{appendix}
\usepackage{multirow}

\begin{document}
\def\qq{\langle \bar q q \rangle}
\def\uu{\langle \bar u u \rangle}
\def\dd{\langle \bar d d \rangle}
\def\sp{\langle \bar s s \rangle}
\def\GG{\langle g_s^2 G^2 \rangle}
\def\Tr{\mbox{Tr}}
\def\figt#1#2#3{
        \begin{figure}
        $\left. \right.$
        \vspace*{-2cm}
        \begin{center}
        \includegraphics[width=10cm]{#1}
        \end{center}
        \vspace*{-0.2cm}
        \caption{#3}
        \label{#2}
        \end{figure}
	}
	
\def\figb#1#2#3{
        \begin{figure}
        $\left. \right.$
        \vspace*{-1cm}
        \begin{center}
        \includegraphics[width=10cm]{#1}
        \end{center}
        \vspace*{-0.2cm}
        \caption{#3}
        \label{#2}
        \end{figure}
                }

\def\ds{\displaystyle}
\def\beq{\begin{equation}}
\def\eeq{\end{equation}}
\def\bea{\begin{eqnarray}}
\def\eea{\end{eqnarray}}
\def\beeq{\begin{eqnarray}}
\def\eeeq{\end{eqnarray}}
\def\ve{\vert}
\def\vel{\left|}
\def\ver{\right|}
\def\nnb{\nonumber}
\def\ga{\left(}
\def\dr{\right)}
\def\aga{\left\{}
\def\adr{\right\}}
\def\lla{\left<}
\def\rra{\right>}
\def\rar{\rightarrow}
\def\lrar{\leftrightarrow}  
\def\nnb{\nonumber}
\def\la{\langle}
\def\ra{\rangle}
\def\ba{\begin{array}}
\def\ea{\end{array}}
\def\tr{\mbox{Tr}}
\def\ssp{{\Sigma^{*+}}}
\def\sso{{\Sigma^{*0}}}
\def\ssm{{\Sigma^{*-}}}
\def\xis0{{\Xi^{*0}}}
\def\xism{{\Xi^{*-}}}
\def\qs{\la \bar s s \ra}
\def\qu{\la \bar u u \ra}
\def\qd{\la \bar d d \ra}
\def\qq{\la \bar q q \ra}
\def\gGgG{\la g^2 G^2 \ra}
\def\q{\gamma_5 \not\!q}
\def\x{\gamma_5 \not\!x}
\def\g5{\gamma_5}
\def\sb{S_Q^{cf}}
\def\sd{S_d^{be}}
\def\su{S_u^{ad}}
\def\sbp{{S}_Q^{'cf}}
\def\sdp{{S}_d^{'be}}
\def\sup{{S}_u^{'ad}}
\def\ssp{{S}_s^{'??}}

\def\sig{\sigma_{\mu \nu} \gamma_5 p^\mu q^\nu}
\def\fo{f_0(\frac{s_0}{M^2})}
\def\ffi{f_1(\frac{s_0}{M^2})}
\def\fii{f_2(\frac{s_0}{M^2})}
\def\O{{\cal O}}
\def\sl{{\Sigma^0 \Lambda}}
\def\es{\!\!\! &=& \!\!\!}
\def\ap{\!\!\! &\approx& \!\!\!}
\def\ar{&+& \!\!\!}
\def\ek{&-& \!\!\!}
\def\kek{\!\!\!&-& \!\!\!}
\def\cp{&\times& \!\!\!}
\def\se{\!\!\! &\simeq& \!\!\!}
\def\eqv{&\equiv& \!\!\!}
\def\kpm{&\pm& \!\!\!}
\def\kmp{&\mp& \!\!\!}
\def\mcdot{\!\cdot\!}
\def\erar{&\rightarrow&}


\def\simlt{\stackrel{<}{{}_\sim}}
\def\simgt{\stackrel{>}{{}_\sim}}

\def\olra{\stackrel{\leftrightarrow}}
\def\ola{\stackrel{\leftarrow}}
\def\ora{\stackrel{\rightarrow}}


\title{
         {\Large
                 {\bf
Magnetic dipole moments of the heavy tensor mesons in QCD
                 }
         }
      }

\author{\vspace{1cm}\\
{\small T. M. Aliev \thanks {e-mail:
taliev@metu.edu.tr}~\footnote{permanent address: Institute of
Physics,Baku,Azerbaijan}\,\,, T. Barakat \thanks {e-mail:
tbarakat@KSU.EDU.SA}\,\,, M. Savc{\i} \thanks
{e-mail: savci@metu.edu.tr}} \\
{\small Physics Department, Middle East Technical University,
06531 Ankara, Turkey }\\
{\small $^\ddag$ Physics and Astronomy Department, King Saud University, Saudi Arabia}}

\date{}

\begin{titlepage}
\maketitle
\thispagestyle{empty}

\begin{abstract}
The magnetic dipole moments of the ${\cal D}_2$,
and ${\cal D}_{S_2}$, ${\cal B}_2$, and ${\cal B}_{S_2}$ heavy tensor mesons
are estimated in framework of the light cone QCD sum rules. It is observed
that the magnetic dipole moments for the charged mesons are larger than that
of its neutral counterpart. It is found that the $SU(3)$ flavor symmetry
violation is about 10\% in both $b$ and $c$ sectors.
\end{abstract}

\vspace{1cm}
~~~PACS number(s): 11.55.Hx, 13.25.Jx, 13.40.Em
\end{titlepage}

\section{Introduction}

Recent years were quite productive in field of the particle spectroscopy.
Many charmonium and bottomonium like states are observed by BaBar, Belle,
LHCb and BES III collaborations \cite{Rnsu01}. These progresses in
experiments stimulated further theoretical studies and experimental
investigations on this subject \cite{Rnsu02}. Considerable progress has also been
made on spectroscopy of the conventional heavy mesons states containing
single charm and bottom quarks such as ${\cal D}_{s_I}(2700)$,
${\cal D}_{s_J}^\ast(2860)$, ${\cal D}_{s_J}(3040)$, ${\cal D}_{J}(2580)$,
${\cal D}_{J}(2740)$, ${\cal D}_{J}^\ast(2760)$, ${\cal D}_{J}(3000)$,
${\cal D}_{J}^\ast(3000)$, ${\cal B}_{1}(5721)$, ${\cal B}_{2}^\ast(5747)$,
${\cal B}_{s_1}(5830)$, ${\cal B}_{s_2}^\ast(5840)$, ${\cal D}(5970)$, etc
\cite{Rnsu03}. Soon after D0 Collaboration observed the ${\cal B}_{1}(5721)$ and
${\cal B}_{2}(5747)$ states \cite{Rnsu04}, which were both confirmed by the
CDF Collaboration \cite{Rnsu05}. The CDF Collaboration further observed the
${\cal D}_{s_1}(5830)$ and  ${\cal B}_{s_2}^\ast(5840)$ states \cite{Rnsu06}
which in turn confirmed by the D0 Collaboration \cite{Rnsu07}. Moreover, the
masses of the ${\cal D}_{s_1}(5830)$ and ${\cal B}_{s_2}^\ast(5840)$ states
were determined more accurately by the LHCb Collaboration \cite{Rnsu08}.     

The masses and decay constants of the heavy tensor meson 
${\cal D}_{2}^\ast(2740)$ and ${\cal D}_{s_2}^\ast(2573)$  
states were first studied within the framework of the QCD sum rules method in
\cite{Rnsu09,Rnsu10}. These mesons, as well as the ${\cal B}_{2}^\ast(5747)$,
${\cal B}_{s_1}(5830)$ and ${\cal B}_{s_2}^\ast(5840)$ tensor mesons have
recently been studied within the same approach in \cite{Rnsu10}. Note that
light tensor mesons without, and with strange quark have also been analyzed 
in framework of the QCD sum rules method in \cite{Rnsu11} and \cite{Rnsu12},
respectively.

One of the most promising approaches in investigating the properties of
mesons and hadrons is the study of the electromagnetic form factors and multipole
moments. These studies can provide useful information about their internal
structures. The electromagnetic properties of usual mesons, as well as
photons and neutrons have comprehensively been studied from theoretical and
experimental sides, and at present it is the subject of the growing interest
from both sides. However the study of the electromagnetic properties of the
tensor mesons has received less interest, and therefore more effort is
needed in this respect. The magnetic moments of the light tensor mesons were
investigated in framework of the light cone QCD sum rules method (LCSR) in
\cite{Rnsu13}.
In the present work we calculate the magnetic dipole moments of the recently
discovered heavy tensor mesons in LCSR.

The paper is organized as follows. In section 2, the light cone QCD sum
rules are constructed for the magnetic dipole moments of the heavy tensor
mesons. In section 3, the numerical analysis is performed for the obtained
sum rules.

\section{Theoretical framework}
Before presenting the light cone sum rules for the magnetic dipole moments
of the heavy tensor mesons, let us first introduce the matrix element which
corresponds to the transition of the heavy tensor meson with momentum $p+q$
to the heavy tensor meson with momentum $p$ in presence of the
electromagnetic field, i.e.,
\bea
\label{ensu01}
&&\lla T_Q(p,\varepsilon) \vel j_\rho^{el} \ver T_Q(p+q,\varepsilon) \rra =
\varepsilon_{\xi\sigma}^\ast(p) \Bigg\{2 p_\rho \Bigg[ g^{\xi\lambda}
g^{\sigma\tau} F_1(Q^2)  - g^{\sigma\tau} {q^\xi q^\lambda \over 2
m_{T_Q}^2} F_3(Q^2) \nnb \\
\ar {q^\xi q^\lambda \over 2 m_{T_Q}^2}\, {q^\sigma q^\tau \over
2 m_{T_Q}^2} F_5(Q^2) \Bigg]
+ \Big(g^{\rho\sigma} q^\tau - g^{\rho\tau} q^\sigma \Big)
\Bigg[ g^{\xi\lambda} F_2(Q^2) - {q^\xi q^\lambda \over 2 m_{T_Q}^2}
F_4(Q^2) \Bigg]
\Bigg\} \varepsilon_{\lambda\tau} (p+q)~,
\eea
where $F_i(Q^2)$ are the form factors, and $T_Q(p,\varepsilon)$ means heavy
tensor meson with momentum $p$ and polarization tensor
$\varepsilon_{\alpha\beta}$. Since we are interested with the magnetic
dipole moments of heavy tensor mesons, the values of the form factors at
$Q^2=-q^2=0$ need to be calculated. The transition under consideration can
be described by the following correlation function,
\bea
\label{ensu02}
\Pi^{\mu\nu\rho\alpha\beta} (p,q) = - \int d^4x \int d^4y e^{i(px+qy)}
\lla 0 \vel {\cal T} \Big\{ j_{\mu\nu} (0) j_\rho^{el}(y) \bar
j_{\alpha\beta}(x) \Big\}
\ver 0 \rra~,
\eea
where $j_{\mu\nu}$ is the interpolating current of the heavy tensor meson,
and $j_\rho^{el}$ is the electromagnetic current given as,
\bea
j_\rho^{el} = e_q \bar{q} \gamma_\rho q + e_Q \bar{Q} \gamma_\rho Q~, \nnb
\eea
with the electric charges $e_q$ and $e_Q$ of the light and heavy quarks,
respectively.
The coupling of the tensor meson current $j_{\mu\nu}$ to the tensor state is
defined as,
\bea
\label{ensu03}
\lla 0 \vel j_{\mu\nu} \ver T_Q(p,\varepsilon) \rra = m_{T_Q}^3 g_{T_Q}
\varepsilon_{\mu\nu}~,
\eea
where $m_{T_Q}$ is the mass, and $g_{T_Q}$ is the coupling constant of the
tensor meson. 

It is more convenient to rewrite the correlator (\ref{ensu02}) by introducing
the electromagnetic background field strength tensor
\bea
F_{\mu\nu} = i (q_\mu\varepsilon_\nu-q_\nu\varepsilon_\mu)
e^{iqx}~,\nnb
\eea
of the plane wave, in the following form
\bea
\label{ensu04}
\Pi_{\mu\nu\rho\alpha\beta} \, \varepsilon^\rho = i \int d^4x e^{ipx}
\lla 0 \vel T \Big\{ j_{\mu\nu} (x) \bar{j}_{\alpha\beta} (0) \ver 0
\rra_F~,
\eea
where $F$ means that all vacuum expectation values are calculated in the
background electromagnetic field.
 The correlation function (\ref{ensu02}) can be obtained by expanding the
correlation function (\ref{ensu04}) in powers of the background field, and
taking into account the terms that are linear in $F_{\mu\nu}$ which
corresponds to the single photon radiation. The advantage of using the
background background field is that it separates the hard and soft photon
radiation in an explicitly gauge invariant way (for more about the details of
the background field method, see \cite{Rnsu14} and \cite{Rnsu15}).  

After these preliminary remarks, we can now proceed deriving the light cone
QCD sum rules for the magnetic dipole moments of the heavy tensor mesons.
These sum rules can
be obtained by calculating the correlator function in terms of mesons (physical
part) from one side; and calculating the same correlation function in terms
of quark-gluon degrees of freedom by using the operator product expansion
(OPE) in deep Eucledian region from theoretical side.  Matching these two
representations of the same correlation
function, the sum rules for the magnetic dipole moments of the heavy tensor
mesons are obtained.

Calculation of the correlation function from the physical
side is performed by inserting the complete set of tensor meson states
having the same quantum number as that of the interpolating current, and
isolating he ground state, as the result of which we get,
\bea
\label{ensu05}
\Pi_{\mu\nu\rho\alpha\beta}\, \varepsilon^\rho  \es 
i \varepsilon^\rho 
{\lla 0 \vel j_{\mu\nu} \ver
T_Q(p,\varepsilon) \rra \over p^2-m_{T_Q}^2} \lla T_Q(p,\varepsilon) \vel
j_\rho^{el} \ver T_Q(p+q,\varepsilon) \rra
{\lla T_Q(p+q,\varepsilon) \vel j_{\alpha\beta}^\dagger \ver
0 \rra \over (p+q)^2-m_{T_Q}^2}+ \cdots~,
\eea
where dots mean the
contribution of the higher states and continuum.

From experimental point of view the multipole form factors are more useful
than the form factors given in Eq. (\ref{ensu01}). The relations between the
two sets of the form factors for any arbitrary $q^2$ are derived in
\cite{Rnsu16}. For the real photon $(q^2=0)$ these relations are given as:
\bea
\label{ensu06}
F_1(0) \es G_{E_0}(0)~, \nnb \\
F_2(0) \es G_{M_1}(0)~, \nnb \\
F_3(0) \es - 2 G_{E_0}(0) +  G_{E_2}(0) + G_{M_1}(0)~, \nnb \\
F_4(0) \es - G_{M_1}(0) + G_{M_3}(0)~, \nnb \\
F_5(0) \es G_{E_0}(0) - [ G_{E_2}(0) + G_{M_1}(0) ] + G_{E_4}(0) +
G_{M_3}(0)~,
\eea
where $G_{E_\ell}(0)$ and $G_{M_\ell}(0)$ are the electric and magnetic
multipoles. 

Calculation of the correlator function from the physical side is performed by
inserting the complete set of tensor meson states in Eqs.
(\ref{ensu01}--\ref{ensu04}), from which for the hadronic part we get,
\bea
\label{ensu07}
\Pi_{\mu\nu\rho\alpha\beta} (p,q) \, \varepsilon^\rho \es
{m_{T_Q}^6 g_{T_Q}^2 \over (p^2-m_{T_Q}^2) [(p+q)^2-m_{T_Q}^2]}
\varepsilon_{\mu\nu}(p) \varepsilon_{\xi\sigma}^\ast(p)
\Bigg\{2 (p\mcdot \varepsilon) \Bigg[ g^{\xi\lambda}
g^{\sigma\tau} F_1(Q^2)  \nnb \\
\ek g^{\sigma\tau} {q^\xi q^\lambda \over 2
m_{T_Q}^2} F_3(Q^2) + 
{q^\xi q^\lambda \over 2 m_{T_Q}^2}\, {q^\sigma q^\tau \over
2 m_{T_Q}^2} F_5(Q^2)
\Bigg]\nnb \\
\ar \Big(\varepsilon^\sigma q^\tau - \varepsilon^\tau q^\sigma \Big)
\Bigg[ g^{\xi\lambda} F_2(Q^2) - {q^\xi q^\lambda \over 2 m_{T_Q}^2}
F_4(Q^2) \Bigg]
\Bigg\} \varepsilon_{\lambda\tau} (p+q) \varepsilon_{\alpha\beta}^\ast (p+q)~.
\eea
Summation over the polarizations of the heavy tensor mesons
can be performed by using the relation,
\bea
\label{ensu08}
\varepsilon_{\rho\sigma}^\ast (p) \varepsilon_{\lambda\tau} (p) \es
{1\over 2} {\cal P}_{\rho\lambda}(p) {\cal P}_{\sigma\tau}(p) +
{1\over 2} {\cal P}_{\rho\tau}(p) {\cal P}_{\sigma\lambda}(p) -
{1\over 3} {\cal P}_{\rho\sigma}(p) {\cal P}_{\lambda\tau}(p)~,
\eea
where
\bea
{\cal P}_{\rho\lambda}(p)  =  \Bigg( - g_{\rho\lambda} + {p_\rho p_\lambda \over
m_{T_Q}^2} \Bigg)~.\nnb
\eea
Using Eqs. (\ref{ensu06}) and (\ref{ensu08}) in Eq. (\ref{ensu07}), for the
physical part of the correlation function we get,
\bea
\label{ensu09}
\Pi_{\mu\nu\rho\alpha\beta} (p,q) \, \varepsilon^\rho \es
{m_{T_Q}^6 g_{T_Q}^2 \over (p^2-m_{T_Q}^2) [(p+q)^2-m_{T_Q}^2]}
\Big\{
{1\over 2} {\cal P}_{\mu\xi}(p) {\cal P}_{\nu\sigma}(p) +
{1\over 2} {\cal P}_{\mu\sigma}(p) {\cal P}_{\nu\xi}(p) -
{1\over 3} {\cal P}_{\mu\nu}(p) {\cal P}_{\xi\sigma}(p)
\Big\} \nnb \\
\cp \Bigg\{2 (p\mcdot \varepsilon) \Bigg[ g^{\xi\lambda}
g^{\sigma\tau} G_{E_0}(0)
- g^{\sigma\tau} {q^\xi q^\lambda \over 2
m_{T_Q}^2} \Big( - 2 G_{E_0}(0)+G_{E_2}(0) + G_{M_1}(0)\Big) \nnb \\
\ar {q^\xi q^\lambda \over 2 m_{T_Q}^2}\, {q^\sigma q^\tau \over
2 m_{T_Q}^2} \Big(G_{E_0}(0) - [G_{E_2}(0) + G_{M_2}(0)] +
G_{E_4}(0) + G_{M_3}(0)\Big)
\Bigg]\nnb \\
\ar \Big(\varepsilon^\sigma q^\tau - \varepsilon^\tau q^\sigma \Big)
\Bigg[ g^{\xi\lambda} G_{M_1}(0) - {q^\xi q^\lambda \over 2 m_{T_Q}^2}
\Big( - G_{M_1}(0) + G_{M_2}(0)\Big) \Bigg]
\Bigg\} \nnb \\
\cp \Big\{
{1\over 2} {\cal P}_{\lambda\alpha}(p+q) {\cal P}_{\tau\beta}(p+q) +
{1\over 2} {\cal P}_{\lambda\beta}(p+q) {\cal P}_{\alpha\tau}(p+q) -
{1\over 3} {\cal P}_{\lambda\tau}(p) {\cal P}_{\alpha\beta}(p+q)
\Big\}~.
\eea

It follows from Eq. (\ref{ensu09}) that the correlation function
$\Pi_{\mu\nu\rho\alpha\beta} \, \varepsilon^\rho$
contains various independent structures, each of
which can be used in determination of the multipole moments of heavy tensor
mesons. In the
present work we restrict ourselves to the calculation of the magnetic dipole
moment $G_{M_1}$, and for this goal we choose the structure
$(\varepsilon^\beta q^\nu - \varepsilon^\nu q^\beta) g_{\mu\alpha}$. This
structure has advantage over the others since it does not contain
contributions coming from the contact terms (for more detail about the
contact terms, see for example \cite{Rnsu17}). As the result we get the
following sum rules for the magnetic dipole moment of the heavy tensor
mesons,
\bea
\label{ensu10}
\Pi_{\mu\nu\rho\alpha\beta} \, \varepsilon^\rho =
g_{\mu\alpha} (\varepsilon^\beta q^\nu -
\varepsilon^\nu q^\beta) 
{m_{T_Q}^6 g_{T_Q}^2 \over (p^2-m_{T_Q}^2) [(p+q)^2 - m_{T_Q}^2]}  
{1\over 4} \Bigg[ G_{M_1} + \mbox{\rm other structures} \Bigg]~.
\eea
Denoting the coefficient of the $(\varepsilon^\beta q^\nu - \varepsilon^\nu
q^\beta) g_{\mu\alpha}$ structure as the invariant function $\Pi$, we get
the following final result for the physical part of the correlator function
for the magnetic dipole moment $G_{M_1}$,
\bea
\label{ensu11}
\Pi = {m_{T_Q}^6 g_{T_Q}^2 \over (p^2-m_{T_Q}^2) [(p+q)^2 - m_{T_Q}^2]}
{1\over 4} G_{M_1}~.
\eea 
In order to construct the corresponding sum rules the correlation function
$\Pi_{\mu\nu\rho\alpha\beta} \, \varepsilon^\rho$
needs to be calculated from the QCD side in terms
of quark and gluon degrees of freedom using the operator product expansion,
for which we need the interpolating current $j_{\mu\nu}$.  The interpolating
current for the ground state heavy tensor meson with the quantum numbers
$2^+$ can be chosen as,
\bea
\label{ensu12}  
j_{\mu\nu} = {i\over 2} \left[ \bar{q}(x) \left( \gamma_\mu \olra{\cal D}_\nu
+ \gamma_\nu \olra{\cal D}_\mu \right) Q(x) \right]~,
\eea
where $q$ and $Q$ denote the light and heavy quark fields, and the
derivative operator $\olra{\cal D}_\mu$ with respect to $x_\mu$ can be written as,
\bea
\olra{\cal D}_\mu (x) = {1\over 2} \Big[
\ora{\cal D}_\mu (x) -
\ola{\cal D}_\mu (x) \Big]~, \nnb
\eea 
which acts on the right and left sides, and the covariant derivatives
in it are defined as

\bea
\label{ensu13}
\ora{\cal D}_\mu (x) \es \ora{\partial}_\mu (x) - i
{g\over 2} \lambda^a A_\mu^a (x) ~, \nnb \\
\ola{\cal D}_\mu (x) \es \ola{\partial}_\mu (x) + i
{g\over 2} \lambda^a A_\mu^a (x)~,
\eea
where $\lambda^a$ are the Gell-Mann matrices, and $A_\mu^a (x)$ is the gluon
field. In the present work we use the Fock-Schwinger gauge in which the
external field $A_\mu^a (x)$ satisfies the condition $x^\mu A_\mu^a (x)=0$.

The correlation function is calculated from the QCD side in deep Eucledian
region $p^2 \to - \infty$, and $(p+q)^2 \to - \infty$ after contracting the
corresponding heavy and light quark fields, as a result of which we get,
\bea
\label{ensu14}   
\Pi_{\mu\nu\rho\alpha\beta} \, \varepsilon^\rho \es
{-i \over 16} \int e^{ip\cdot x} e^{-i(p+q)\cdot y} d^4x 
\lla 0 \vel \Big\{
S_{q} (y-x) \gamma_\mu \Big[ 
 \ora{\partial}_\nu (x) \ora{\partial}_\beta (y) \right. \right.\nnb \\
\ek \left. \left. 
 \ora{\partial}_\nu (x) \ola{\partial}_\beta (y) -
 \ola{\partial}_\nu (x) \ora{\partial}_\beta (y)  +
 \ola{\partial}_\nu (x) \ola{\partial}_\beta (y) \Big]
S_{Q} (x-y) \gamma_\alpha \Big\} \right. \right.\nnb \\
\ar \left. \left.
\{ \beta \leftrightarrow \alpha \} + \{ \nu \leftrightarrow \mu \} +
\{ \beta \leftrightarrow \alpha,~ \nu \leftrightarrow \mu \}
\phantom{\Big]}\ver 0 \rra_F~,
\eea
where we set $y=0$ after performing derivative with respect to $y$.

We see from Eq. (\ref{ensu14}) that, the expressions of the propagators of
the light and heavy quarks are needed in order to calculate the correlation
function. The light quark propagator in presence of the external field is
calculated in \cite{Rnsu18}, whose expression is given as:

\bea
\label{ensu15}
S_q(x-y) \es S_q^{free} (x-y) - {\qq \over 12} \Bigg[1 -i {m_q \over
4} (\not\!{x}-\not\!{y}) \Bigg] + {(x-y)^2 \over 192} m_0^2 \qq
\Bigg[1 - i {m_q \over 6} (\not\!{x}-\not\!{y}) \Bigg] \nnb \\
\ek i g_s \int_0^1 du \Bigg\{
{\not\!x - \not\!{y} \over 16 \pi^2 (x-y)^2} G_{\mu\nu}(u(x-y)) \sigma^{\mu\nu}
- u (x^\mu-y^\mu) G_{\mu\nu}(u(x-y)) \gamma^\nu \nnb \\
\cp {i \over 4 \pi^2 (x-y)^2}
- i {m_q \over 32 \pi^2} G_{\mu\nu}(u(x-y)) \sigma^{\mu\nu} \Bigg[
\ln\left( -{(x-y)^2
\Lambda^2 \over u} + 2 \gamma_E \right) \Bigg] \Bigg\}~,
\eea
where $\Lambda$ is the scale parameter separating the perturbative and
nonperturbative domains. This parameter is estimated in \cite{Rnsu19} to
have the value $\Lambda=(0.5 \div 1.0)~GeV$; and $S^{free}(x-y)$ is the free quark
operator whose expression is given as:
\bea
\label{ensu16}
S_q^{free} (x-y) = {i (\not\!x-\not\!y) \over 2 \pi^2 (x-y)^4} - {m_q \over 4 \pi^2
(x-y)^2}~.
\eea

The propagator for the heavy quark have the following form in coordinate
space:
\bea
\label{ensu17}
S_Q(x) \es S_Q^{free} -
{g_s \over 16 \pi^2} \int_0^1 du
G_{\mu\nu}(u(x-y)) \Bigg( i \Big[\sigma^{\mu\nu} (\not\!{x}-\not\!{y}) +
(\not\!{x}-\not\!{y})
\sigma^{\mu\nu}\Big] {K_1 (m_Q\sqrt{-(x-y)^2})\over \sqrt{-(x-y)^2}}\nnb \\
\ar 2 \sigma^{\mu\nu} K_0(m_Q\sqrt{-(x-y)^2})\Bigg)\Bigg\} +\cdots~,
\eea
where where $K_i(m_Q\sqrt{-x^2})$ are the modified Bessel functions, and
\bea
S_Q^{free} = {m_Q^2 \over 4 \pi^2} \Bigg\{ {K_1(m_Q\sqrt{-(x-y)^2}) \over \sqrt{-(x-y)^2}} +
i {(\not\!{x}-\not\!{y}) \over -(x-y)^2} K_2(m_Q\sqrt{-(x-y)^2}) \Bigg\}~. \nnb	
\eea
Few words about the expression of the quark propagator are in order. The
complete light cone expansion of the light quark propagator in presence of
the external field is calculated in \cite{Rnsu18}, which includes the
contributions coming from nonlocal three $\bar{q}G q$, and
four-particle $\bar{q}q\bar{q}q$, $\bar{q}G^2 q$ operators. Using the
expansion in conformal spin, one can show that aforementioned contributions
are small (for more detail see \cite{Rnsu20}), therefore we shall neglect them in
further analysis.

There are three type of contributions to the correlation function: 1)
Perturbative part, when photon interacts perturbatively with the quark
propagator (light or heavy). 2) ``Mixed" contributions, which take place
when heavy quark propagator interacts with the photon field perturbatively,
and light quark fields form quark condensate. 3) ``Long distance "
contribution. It takes place when photon is radiated at long distance.

The perturbative contribution is calculated from Eq. (\ref{ensu14}) by
replacing heavy or light propagator with,
\bea                                                                  
\label{ensu18}
S_{\mu\nu}^{ab}(x-y) \to -{1\over 4} \bar{q}^a (x) \Gamma_\rho q^b (y)
\left(\Gamma_\rho \right)_{\mu\nu},
\eea 
where $\Gamma_\rho$ are the full set of Dirac matrices. As has already been
noted, when a photon interacts with the light quark fields matrix elements
of the nonlocal operators such as $\bar{q}(x) \Gamma q(y)$ and $\bar{q}(x)
G_{\mu\nu} \Gamma q(y)$ appear between vacuum and photon states. These
matrix elements are parametrized in terms of the photon distribution
amplitudes (DAs), which are the key nonperturbative parameters in light cone
sum rules, whose explicit expressions are given below,

\bea
\label{ensu19}
&&\langle \gamma(q) \vert  \bar q(x) \sigma_{\mu \nu} q(0) \vert  0
\rangle  = -i e_q \qq (\varepsilon_\mu q_\nu - \varepsilon_\nu
q_\mu) \int_0^1 du e^{i \bar u qx} \left(\chi \varphi_\gamma(u) +
\frac{x^2}{16} \mathbb{A}  (u) \right) \nnb \\ &&
-\frac{i}{2(qx)}  e_q \qq \left[x_\nu \left(\varepsilon_\mu - q_\mu
\frac{\varepsilon x}{qx}\right) - x_\mu \left(\varepsilon_\nu -
q_\nu \frac{\varepsilon x}{q x}\right) \right] \int_0^1 du e^{i \bar
u q x} h_\gamma(u)
\nnb \\
&&\langle \gamma(q) \vert  \bar q(x) \gamma_\mu q(0) \vert 0 \rangle
= e_q f_{3 \gamma} \left(\varepsilon_\mu - q_\mu \frac{\varepsilon
x}{q x} \right) \int_0^1 du e^{i \bar u q x} \psi^v(u)
\nnb \\
&&\langle \gamma(q) \vert \bar q(x) \gamma_\mu \gamma_5 q(0) \vert 0
\rangle  = - \frac{1}{4} e_q f_{3 \gamma} \epsilon_{\mu \nu \alpha
\beta } \varepsilon^\nu q^\alpha x^\beta \int_0^1 du e^{i \bar u q
x} \psi^a(u)
\nnb \\
&&\langle \gamma(q) | \bar q(x) g_s G_{\mu \nu} (v x) q(0) \vert 0
\rangle = -i e_q \qq \left(\varepsilon_\mu q_\nu - \varepsilon_\nu
q_\mu \right) \int {\cal D}\alpha_i e^{i (\alpha_{\bar q} + v
\alpha_g) q x} {\cal S}(\alpha_i)
\nnb \\
&&\langle \gamma(q) | \bar q(x) g_s \tilde G_{\mu \nu} i \gamma_5 (v
x) q(0) \vert 0 \rangle = -i e_q \qq \left(\varepsilon_\mu q_\nu -
\varepsilon_\nu q_\mu \right) \int {\cal D}\alpha_i e^{i
(\alpha_{\bar q} + v \alpha_g) q x} \tilde {\cal S}(\alpha_i)
\nnb \\
&&\langle \gamma(q) \vert \bar q(x) g_s \tilde G_{\mu \nu}(v x)
\gamma_\alpha \gamma_5 q(0) \vert 0 \rangle = e_q f_{3 \gamma}
q_\alpha (\varepsilon_\mu q_\nu - \varepsilon_\nu q_\mu) \int {\cal
D}\alpha_i e^{i (\alpha_{\bar q} + v \alpha_g) q x} {\cal
A}(\alpha_i)
\nnb \\
&&\langle \gamma(q) \vert \bar q(x) g_s G_{\mu \nu}(v x) i
\gamma_\alpha q(0) \vert 0 \rangle = e_q f_{3 \gamma} q_\alpha
(\varepsilon_\mu q_\nu - \varepsilon_\nu q_\mu) \int {\cal
D}\alpha_i e^{i (\alpha_{\bar q} + v \alpha_g) q x} {\cal
V}(\alpha_i) \nnb \\ && \langle \gamma(q) \vert \bar q(x)
\sigma_{\alpha \beta} g_s G_{\mu \nu}(v x) q(0) \vert 0 \rangle  =
e_q \qq \left\{
        \left[\left(\varepsilon_\mu - q_\mu \frac{\varepsilon x}{q x}\right)\left(g_{\alpha \nu} -
        \frac{1}{qx} (q_\alpha x_\nu + q_\nu x_\alpha)\right) \right. \right. q_\beta
\nnb \\ && -
         \left(\varepsilon_\mu - q_\mu \frac{\varepsilon x}{q x}\right)\left(g_{\beta \nu} -
        \frac{1}{qx} (q_\beta x_\nu + q_\nu x_\beta)\right) q_\alpha
\nnb \\ && -
         \left(\varepsilon_\nu - q_\nu \frac{\varepsilon x}{q x}\right)\left(g_{\alpha \mu} -
        \frac{1}{qx} (q_\alpha x_\mu + q_\mu x_\alpha)\right) q_\beta
\nnb \\ &&+
         \left. \left(\varepsilon_\nu - q_\nu \frac{\varepsilon x}{q.x}\right)\left( g_{\beta \mu} -
        \frac{1}{qx} (q_\beta x_\mu + q_\mu x_\beta)\right) q_\alpha \right]
   \int {\cal D}\alpha_i e^{i (\alpha_{\bar q} + v \alpha_g) qx} {\cal T}_1(\alpha_i)
\nnb \\ &&+
        \left[\left(\varepsilon_\alpha - q_\alpha \frac{\varepsilon x}{qx}\right)
        \left(g_{\mu \beta} - \frac{1}{qx}(q_\mu x_\beta + q_\beta x_\mu)\right) \right. q_\nu
\nnb \\ &&-
         \left(\varepsilon_\alpha - q_\alpha \frac{\varepsilon x}{qx}\right)
        \left(g_{\nu \beta} - \frac{1}{qx}(q_\nu x_\beta + q_\beta x_\nu)\right)  q_\mu
\nnb \\ && -
         \left(\varepsilon_\beta - q_\beta \frac{\varepsilon x}{qx}\right)
        \left(g_{\mu \alpha} - \frac{1}{qx}(q_\mu x_\alpha + q_\alpha x_\mu)\right) q_\nu
\nnb \\ &&+
         \left. \left(\varepsilon_\beta - q_\beta \frac{\varepsilon x}{qx}\right)
        \left(g_{\nu \alpha} - \frac{1}{qx}(q_\nu x_\alpha + q_\alpha x_\nu) \right) q_\mu
        \right]
    \int {\cal D} \alpha_i e^{i (\alpha_{\bar q} + v \alpha_g) qx} {\cal T}_2(\alpha_i)
\nnb \\ &&+
        \frac{1}{qx} (q_\mu x_\nu - q_\nu x_\mu)
        (\varepsilon_\alpha q_\beta - \varepsilon_\beta q_\alpha)
    \int {\cal D} \alpha_i e^{i (\alpha_{\bar q} + v \alpha_g) qx} {\cal T}_3(\alpha_i)
\nnb \\ &&+
        \left. \frac{1}{qx} (q_\alpha x_\beta - q_\beta x_\alpha)
        (\varepsilon_\mu q_\nu - \varepsilon_\nu q_\mu)
    \int {\cal D} \alpha_i e^{i (\alpha_{\bar q} + v \alpha_g) qx} {\cal T}_4(\alpha_i)
                        \right\}~,
\eea
where $\varphi_\gamma(u)$ is the leading twist-2, $\psi^v(u)$,
$\psi^a(u)$, ${\cal A}$ and ${\cal V}$ are the twist-3, and
$h_\gamma(u)$, $\mathbb{A}$, ${\cal T}_i$ ($i=1,~2,~3,~4$) are the
twist-4 photon DAs, and $\chi$ is the magnetic susceptibility. The photon
DAs are calculated in \cite{Rnsu15} and their explicit expressions are
given in Appendix A.
The measure ${\cal D} \alpha_i$ is defined as
\bea
\int {\cal D} \alpha_i = \int_0^1 d \alpha_{\bar q} \int_0^1 d
\alpha_q \int_0^1 d \alpha_g \delta(1-\alpha_{\bar
q}-\alpha_q-\alpha_g)~.\nnb
\eea

Separating out the coefficient of the structure 
$(\varepsilon^\beta q^\nu - \varepsilon^\nu q^\beta) g_{\mu\alpha}$ from the
QCD and the phenomenological parts of the correlation function and equating
them, we get the magnetic moments of the heavy tensor mesons. In order to
suppress the contributions of the higher states and
continuum, we apply double Borel transformation with respect to the
variables $p^2$ and $(p+q)^2$. As the result of these calculations we obtain
the following sum rules for the magnetic dipole moment of the heavy tensor
mesons,
\bea
\label{ensu20}  
&&{m_{T_Q}^6 g_{T_Q}^2 \over 4} e^{-m_{T_Q}^2/M^2} G_{M_1}(q^2=0) = \nnb \\
\ek{1 \over 1152 \pi^2} 
\Big[e_u \GG M^2 \left(-2 m_b^2 {\cal I}_2 + m_b^4 {\cal I}_3\right)\Big]
- {e^{-m_b^2/M^2}\over 3456 m_b \pi^2} 
 M^2 \Big\{9 m_b \left(e_u \GG + 96 e_b m_b \pi^2 \uu\right) \nnb \\
\ar 4 e_u \pi^2 \Big[18 m_b^2 \uu \left(\mathbb{A} (u_0) + 2
\widetilde{j}_1(h_\gamma) + 4 \widetilde{j}_2(h_\gamma)\right) +
\GG \uu \chi \varphi_\gamma (u_0) + 36 f_{3\gamma} m_b^3 \psi^a (u_0)\Big]\Big\} \nnb \\
\ar {1\over 32 \pi^2} e_b m_b^4 M^4 \left({\cal I}_2 - m_b^2 {\cal I}_3\right) +
{e^{-m_b^2/M^2}\over 96}
e_u M^4 \Big\{8 f_{3\gamma} \widetilde{j}_1(\psi^v) + 8 m_b \uu
\chi \varphi_\gamma (u_0) \nnb \\
\ek f_{3\gamma} \Big[6 \psi^a (u_0) - 4 \psi^v(u_0) +
\psi^{a\prime}(u_0)\Big]\Big\} +
{e^{-m_b^2/M^2}\over 16 \pi^2} e_u M^6 \nnb \\
\ek {1\over 32\pi^2}
m_b^2 M^6 \Big[-2 e_b {\cal I}_2 + 3 e_b m_b^2 {\cal I}_3 + 4 e_b m_b^4 {\cal I}_4 + 
2 e_u m_b^4 {\cal I}_4 + 2 (e_b - e_u) m_b^6 {\cal I}_5 \Big] \nnb \\
\ar {e^{-m_b^2/M^2}\over 6912 M^2}
m_b \Big\{432 e_b m_0^2 m_b^2 \uu + e_u \GG \Big[-4 \uu \mathbb{A} (u_0) + 
     4 (5 - 4 u) \uu \widetilde{j}_1(h_\gamma) \nnb \\
\ar 40 \uu \widetilde{j}_2(h_\gamma) + 
     m_b \Big(-8 m_b \uu \chi \varphi_\gamma (u_0) + f_{3\gamma} (-8 \widetilde{j}_1(\psi^v) + 
         6 \psi^a (u_0) - 4 \psi^v(u_0) + \psi^{a\prime}(u_0))\Big)\Big]\Big\} \nnb \\
\ar {e^{-m_b^2/M^2}\over 3456 M^4}
e_u \GG m_b^3 \Big[\uu \Big(\mathbb{A} (u_0) + 2 \widetilde{j}_1(h_\gamma) + 4
\widetilde{j}_2(h_\gamma)\Big) + 
   2 f_{3\gamma} m_b \psi^a (u_0) \Big] \nnb \\
\ar {e^{-m_b^2/M^2}\over 3456 M^6}
e_u \GG m_b^5 \uu \mathbb{A} (u_0)
+ {e^{-m_b^2/M^2}\over 3456 m_b \pi^2}
e_u \Big\{4 (\GG - 18 m_b^4) \pi^2 \uu \mathbb{A} (u_0) \nnb \\
\ar \GG \Big[-3 m_b^3 + \pi^2 \Big(4 (2 + u) \uu \widetilde{j}_1(h_\gamma) + 
16 \uu \widetilde{j}_2(h_\gamma) + m_b \{-12 m_b \uu \chi \varphi_\gamma (u_0) \nnb \\ 
\ar f_{3\gamma} [-12 \widetilde{j}_1(\psi^v) - 2 \psi^a (u_0) + 
           (-2 + u) (4 \psi^v(u_0) - \psi^{a\prime}(u_0))]\}\Big)\Big]
\Big\}~,
\eea
where
\bea
u_0={M_1^2 \over M_1^2 +M_2^2}~,~~~~~M^2={M_1^2 M_2^2 \over M_1^2 +M_2^2}~.\nnb
\eea

The functions $i_n~(n=1,2)$, and $\widetilde{j}_1(f(u))$
are defined as:
\bea
\widetilde{j}_1(f(u)) \es \int_{u_0}^1 du f(u)~, \nnb \\
\widetilde{j}_2(f(u)) \es \int_{u_0}^1 du (u-u_0) f(u)~, \nnb \\
{\cal I}_n \es \int_{m_b^2}^{s_0} ds\, {e^{-s/M^2} \over s^n}~,\nnb
\eea
where $s_0$ is the continuum threshold.


Since we
have the same heavy tensor mesons in the initial and final states, we can
set $M_1^2=M_2^2=2 M^2$, as the result of which we have,
\bea
u_0={M_1^2 \over (M_1^2+M_2^2)}={1\over 2}~.\nnb
\eea
Physically this means that each quark and antiquark carries the half the
photon  momentum.

\section{Numerical analysis}

This section is devoted to the numerical analysis of the sum rules for the
magnetic dipole moments of the heavy tensor mesons obtained in the previous
section. The values of the input parameters entering into sum rules are,
$\uu(\mu=1~GeV) = \dd(\mu=1~GeV)  = -(0.243)^3~GeV^3$,
$\sp \ve_{\mu=1~GeV} = (0.8 \pm 0.2) \uu(\mu=1~GeV)$,
$m_0^2=(0.8\pm 0.2)~GeV^2$ which are obtained from the mass sum rule
analysis for the light baryons \cite{Rnsu21,Rnsu22}, and $B$, $B^\ast$
mesons \cite{Rnsu23}. For the heavy quark masses we have
used their $\overline{MS}$ values, which are given as:
$\bar{m}_b(\bar{m}_b)=(4.16\pm 0.03)~GeV$ and $\bar{m}_c(\bar{m}_c)=(1.28\pm
0.03)~GeV$ \cite{Rnsu24,Rnsu25}. The magnetic susceptibility of quarks was
estimated in \cite{Rnsu26,Rnsu27,Rnsu28} in framework of the QCD sum rules.
As we have noted earlier, the residues and masses
of the heavy tensor mesons were calculated within the QCD sum rules method
in \cite{Rnsu09,Rnsu10}, and their values are $g_{{\cal D}_2} = 0.0228  \pm
0.0068$, $g_{{\cal D}_{S_2}} = 0.023 \pm 0.011$,
$g_{{\cal B}_2} = 0.0050 \pm 0.0005$, $g_{{\cal B}_{S_2}} =
0.0060 \pm 0.0005$.

Having decided the values of the input parameters, we are ready now to perform
numerical analysis of the sum rules for the magnetic dipole moment of the
heavy tensor mesons.The sum rule contains two unphysical parameters: a) The
Borel mass parameter $M^2$, and continuum threshold $s_0$. It is known
that the physical results should be independent of the these parameters.
Therefore, our primary goal is to find such domain of these parameters for
which the magnetic dipole moment is practically independent of them. The
``working region" of $M^2$ is determined as follows: The upper bound of
$M^2$ can be found by requiring that the contributions coming from higher
states constitutes about 40\% of the perturbative part. The lower bound of
$M^2$ could be fixed by demanding that the higher twist contributions are
less than that of the leading twist contributions. In other words, the light
cone expansion with increasing twist should be convergent. These
requirements leads us to the following domains for the Borel mass parameter:
\bea
&& 2.0~GeV^2 \le M^2 \le 4.0~GeV^2, \mbox{ for ${\cal D}_2$ and ${\cal
D}_{S_2}$ mesons}~,\nnb \\
&& 4.5~GeV^2 \le M^2 \le 7.0~GeV^2, \mbox{ for ${\cal B}_2$ and ${\cal       
D}_{B_2}$ mesons}~.\nnb
\eea

The second parameter entering to the sum rules is the continuum threshold
$s_0$. Generally speaking, this parameter is not arbitrary and it is related  
to the energy of the first excited state. The energy necessary for the
transition of the meson from ground state to first excited state is
$\sqrt{s_0}-m_{ground}$. This difference varies, usually, from $0.3~GeV$ to
$0.8~GeV$, where in our numerical analysis we have used their average value,
i.e., $\sqrt{s_0}-m_{ground}=0.5~GeV$.

In Figs. (1) ,(2) and (3) we present the dependence of the magnetic dipole
moments of ${\cal D}_2^+$, ${\cal D}_2^0$ and ${\cal D}_{S_2}$ tensor mesons
on $M^2$, at various fixed values of the continuum threshold. From these
figures we get the following results,
\bea
G_{M_1}(q^2=0) = \left\{ \begin{array}{l}
\phantom{-}0.75 \pm 0.25,~\mbox{for}~{\cal D}_2^0 \\
-2.10 \pm 0.20,~\mbox{for}~{\cal D}_2^-\\
-2.20 \pm 0.20,~\mbox{for}~{\cal D}_{S_2}^-
\end{array} \nnb \right. 
\eea
in units of $e/2 m_{T_Q}$.

Similar analysis performed for heavy tensor mesons containing $b$-quark,
whose results are presented in Figs. (4)-(6), respectively. 
From the analysis of these figures we obtain,
\bea
G_{M_1}(q^2=0) = \left\{ \begin{array}{l}
\phantom{-}3.8 \pm 0.7,~\mbox{for}~{\cal B}_2^+ \\
-1.3 \pm 0.3,~\mbox{for}~{\cal B}_2^0\\
-1.4 \pm 0.3,~\mbox{for}~{\cal B}_{S_2}^0
\end{array} \nnb \right.
\eea
in units of $e/2 m_{T_Q}$.

In conclusion, The magnetic dipole moments of the heavy tensor mesons are
calculated in framework of the light cone QCD sum rules. It is observed that
the magnetic moments of the charged tensor mesons are larger compared to the
neutral ones. The $SU(3)$ symmetry breaking in heavy tensor mesons containing
beauty quarks is about 10\%, while in the charmed meson sector it is quite
large.


\newpage


\section*{Appendix A: Photon distribution amplitudes}
\setcounter{equation}{0}
\setcounter{section}{0}


Explicit forms of the photon DAs
\cite{Rnsu15}:

\bea
\varphi_\gamma(u) \es 6 u \bar u \Big[ 1 + \varphi_2(\mu)
C_2^{\frac{3}{2}}(u - \bar u) \Big]~,
\nnb \\
\psi^v(u) \es 3 [3 (2 u - 1)^2 -1 ]+\frac{3}{64} (15
w^V_\gamma - 5 w^A_\gamma)
                        [3 - 30 (2 u - 1)^2 + 35 (2 u -1)^4]~,
\nnb \\
\psi^a(u) \es [1- (2 u -1)^2] [ 5 (2 u -1)^2 -1 ]
\frac{5}{2}
    \Bigg(1 + \frac{9}{16} w^V_\gamma - \frac{3}{16} w^A_\gamma
    \Bigg)~,
\nnb \\
{\cal A}(\alpha_i) \es 360 \alpha_q \alpha_{\bar q} \alpha_g^2
        \Bigg[ 1 + w^A_\gamma \frac{1}{2} (7 \alpha_g - 3)\Bigg]~,
\nnb \\
{\cal V}(\alpha_i) \es 540 w^V_\gamma (\alpha_q - \alpha_{\bar q})
\alpha_q \alpha_{\bar q}
                \alpha_g^2~,
\nnb \\
h_\gamma(u) \es - 10 (1 + 2 \kappa^+ ) C_2^{\frac{1}{2}}(u
- \bar u)~,
\nnb \\
\mathbb{A}(u) \es 40 u^2 \bar u^2 (3 \kappa - \kappa^+ +1 ) +
        8 (\zeta_2^+ - 3 \zeta_2) [u \bar u (2 + 13 u \bar u) + 
                2 u^3 (10 -15 u + 6 u^2) \ln(u) \nnb \\ 
\ar 2 \bar u^3 (10 - 15 \bar u + 6 \bar u^2)
        \ln(\bar u) ]~,
\nnb \\
{\cal T}_1(\alpha_i) \es -120 (3 \zeta_2 + \zeta_2^+)(\alpha_{\bar
q} - \alpha_q)
        \alpha_{\bar q} \alpha_q \alpha_g~,
\nnb \\
{\cal T}_2(\alpha_i) \es 30 \alpha_g^2 (\alpha_{\bar q} - \alpha_q)
    [(\kappa - \kappa^+) + (\zeta_1 - \zeta_1^+)(1 - 2\alpha_g) +
    \zeta_2 (3 - 4 \alpha_g)]~,
\nnb \\
{\cal T}_3(\alpha_i) \es - 120 (3 \zeta_2 - \zeta_2^+)(\alpha_{\bar
q} -\alpha_q)
        \alpha_{\bar q} \alpha_q \alpha_g~,
\nnb \\
{\cal T}_4(\alpha_i) \es 30 \alpha_g^2 (\alpha_{\bar q} - \alpha_q)
    [(\kappa + \kappa^+) + (\zeta_1 + \zeta_1^+)(1 - 2\alpha_g) +
    \zeta_2 (3 - 4 \alpha_g)]~,\nnb \\
{\cal S}(\alpha_i) \es 30\alpha_g^2\{(\kappa +
\kappa^+)(1-\alpha_g)+(\zeta_1 + \zeta_1^+)(1 - \alpha_g)(1 -
2\alpha_g)\nnb \\ 
\ar\zeta_2
[3 (\alpha_{\bar q} - \alpha_q)^2-\alpha_g(1 - \alpha_g)]\}~,\nnb \\
\widetilde {\cal S}(\alpha_i) \es-30\alpha_g^2\{(\kappa -
\kappa^+)(1-\alpha_g)+(\zeta_1 - \zeta_1^+)(1 - \alpha_g)(1 -
2\alpha_g)\nnb \\ 
\ar\zeta_2 [3 (\alpha_{\bar q} -
\alpha_q)^2-\alpha_g(1 - \alpha_g)]\}. \nnb
\eea
The parameters entering  the above DA's are borrowed from
\cite{Rnsu15} whose values are $\varphi_2(1~GeV) = 0$, 
$w^V_\gamma = 3.8 \pm 1.8$, $w^A_\gamma = -2.1 \pm 1.0$, 
$\kappa = 0.2$, $\kappa^+ = 0$, $\zeta_1 = 0.4$, $\zeta_2 = 0.3$, 
$\zeta_1^+ = 0$, and $\zeta_2^+ = 0$.


\newpage

\section*{Figure captions}
{\bf Fig. (1)} The dependence of the magnetic dipole moment of the ${\cal
D}_2^0$ tensor meson on $M^2$, at two fixed values of $s_0=9.0~GeV^2$, and
$s_0=9.5~GeV^2$.\\\\
{\bf Fig. (2)} The same as Fig. (1), but for the ${\cal D}_2^-$ tensor
meson.\\\\
{\bf Fig. (3)} The same as Fig. (1), but for the ${\cal D}_{S_2}^-$ tensor
meson.\\\\
{\bf Fig. (4)} The same as Fig. (1), but for the ${\cal B}_2^+$ tensor meson, at
two fixed values of $s_0=37.5~GeV^2$, and $s_0=40.0~GeV^2$.\\\\
{\bf Fig. (5)} The same as Fig. (4), but for the ${\cal B}_2^0$ tensor
meson.\\\\
{\bf Fig. (6)} The same as Fig. (4), but for the ${\cal B}_{S2}^0$ tensor
meson.


\newpage

\begin{figure}
\vskip 3. cm
    \includegraphics{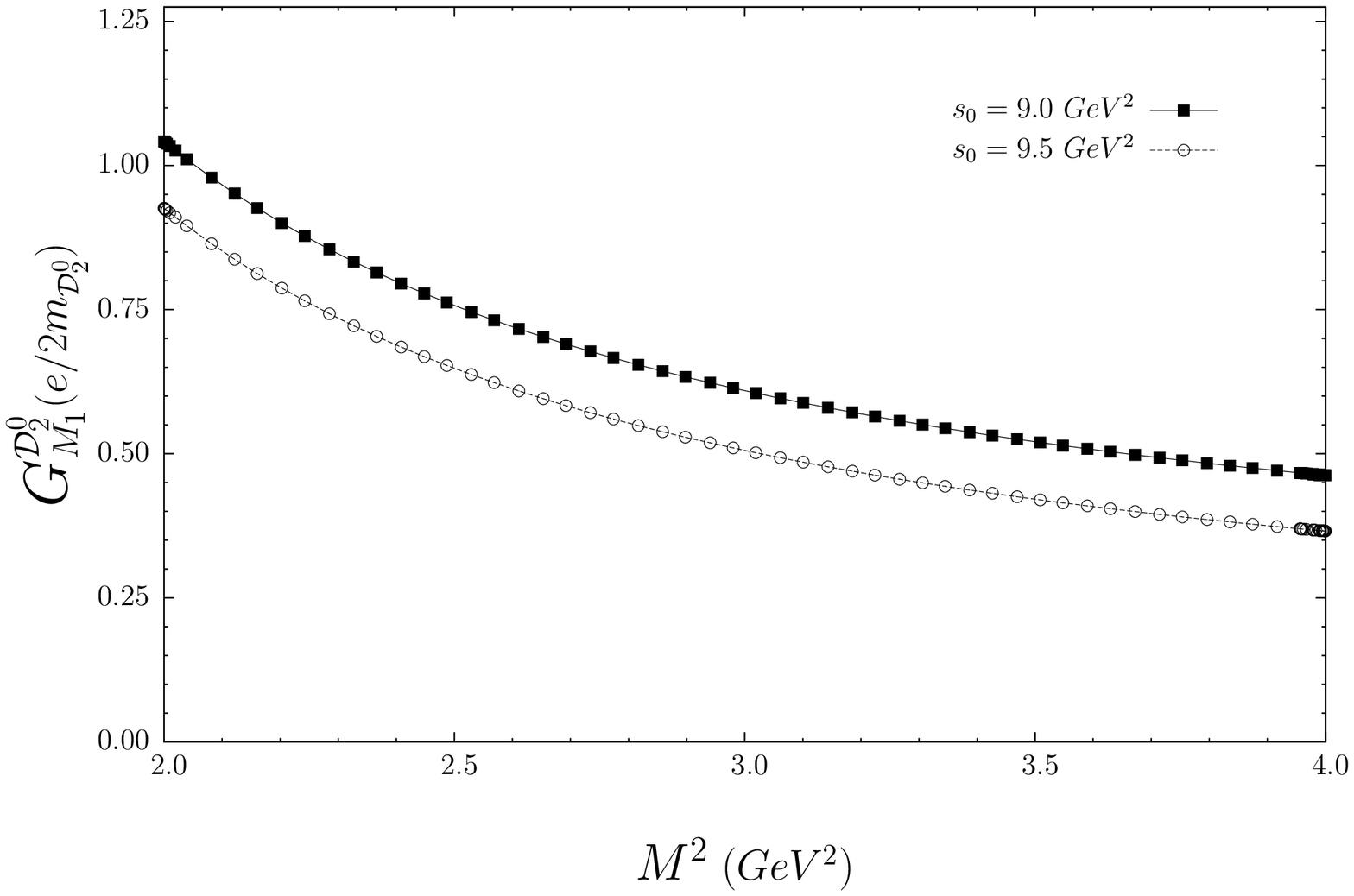}
\vskip 7.0cm
\caption{}
\end{figure}

\begin{figure}
\vskip 3. cm
    \includegraphics{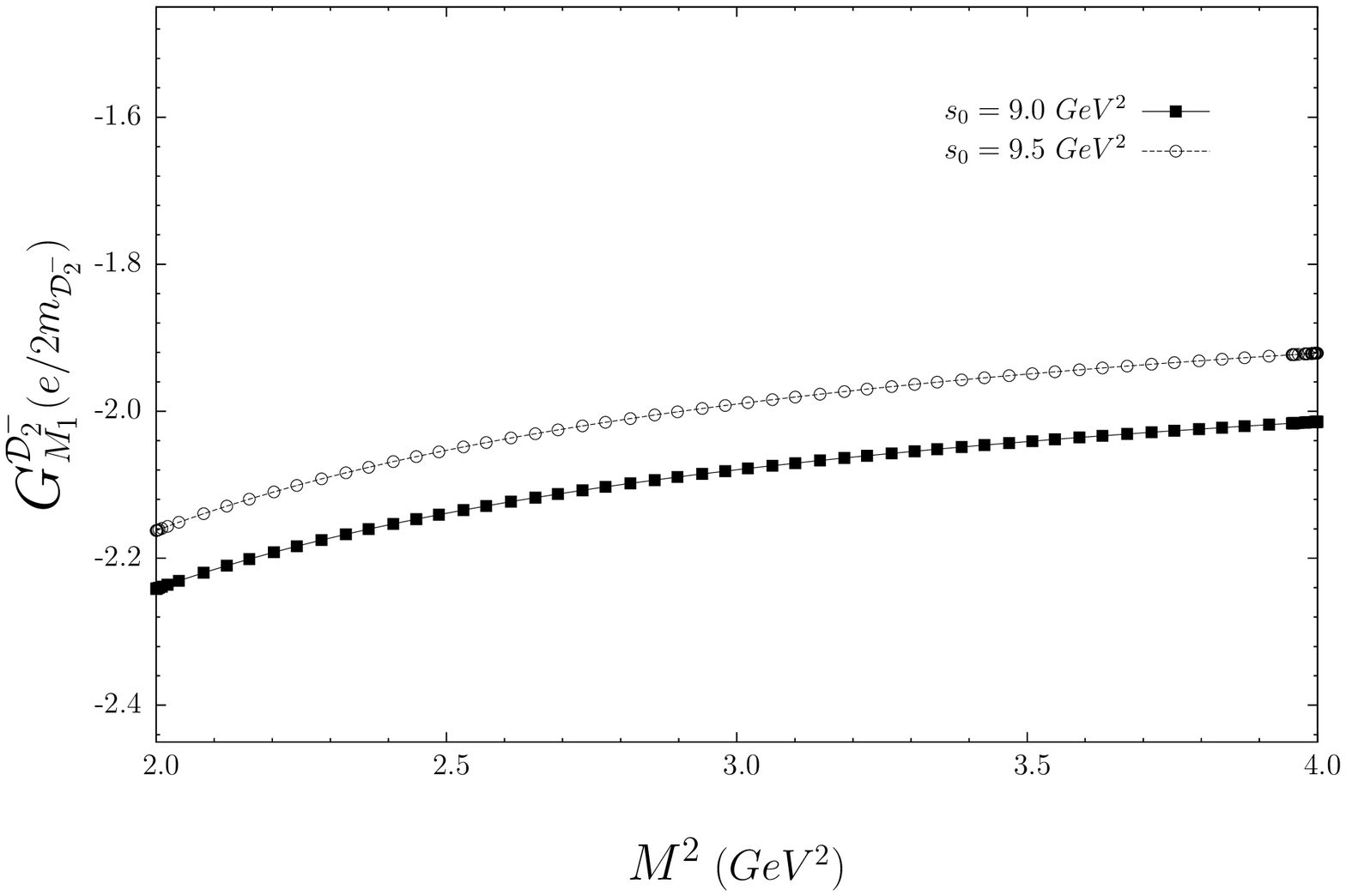}
\vskip 7.0cm
\caption{}
\end{figure}

\begin{figure}
\vskip 3. cm
    \includegraphics{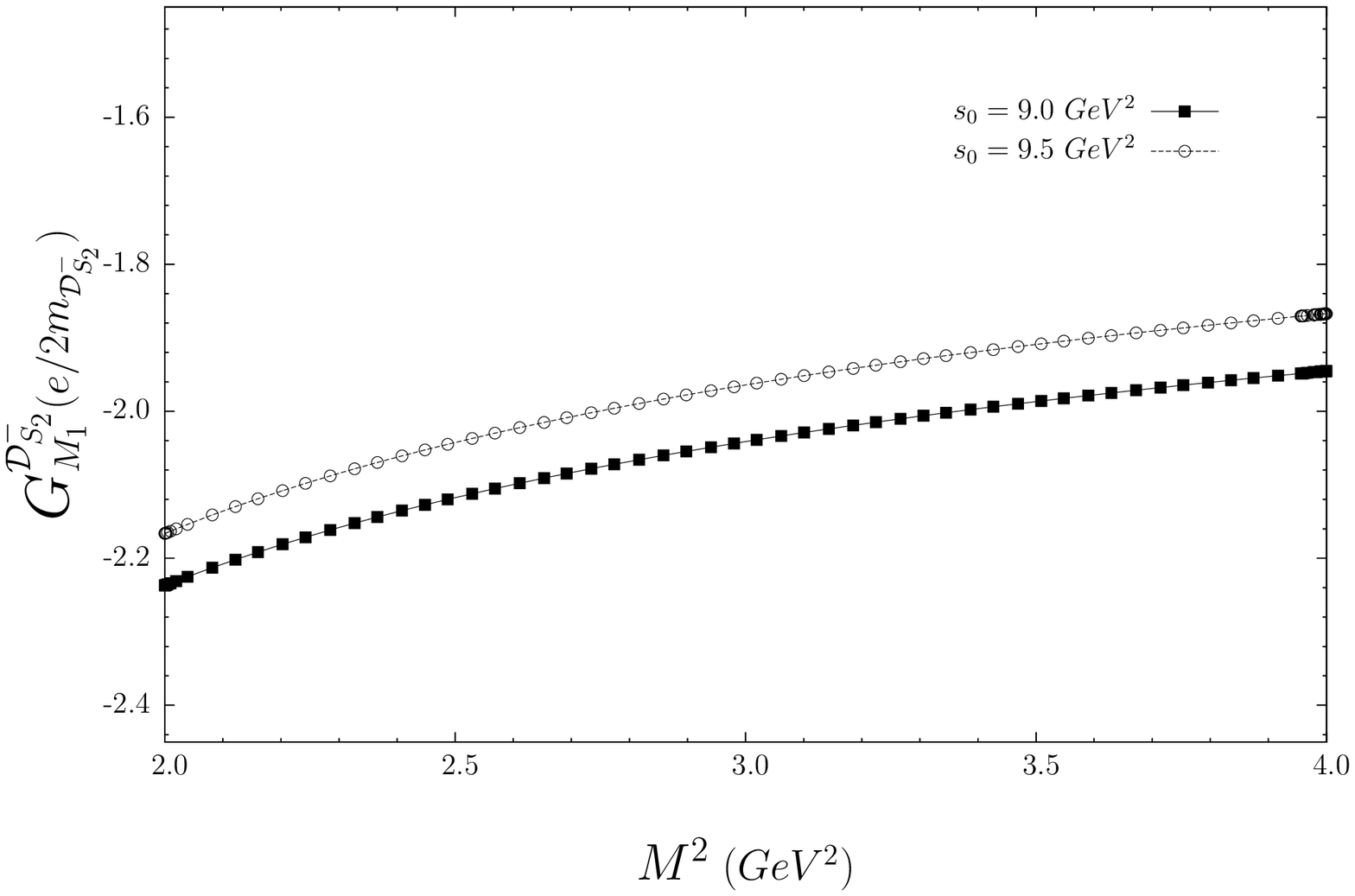}
\vskip 7.0cm
\caption{}
\end{figure}

\begin{figure}
\vskip 3. cm
    \includegraphics{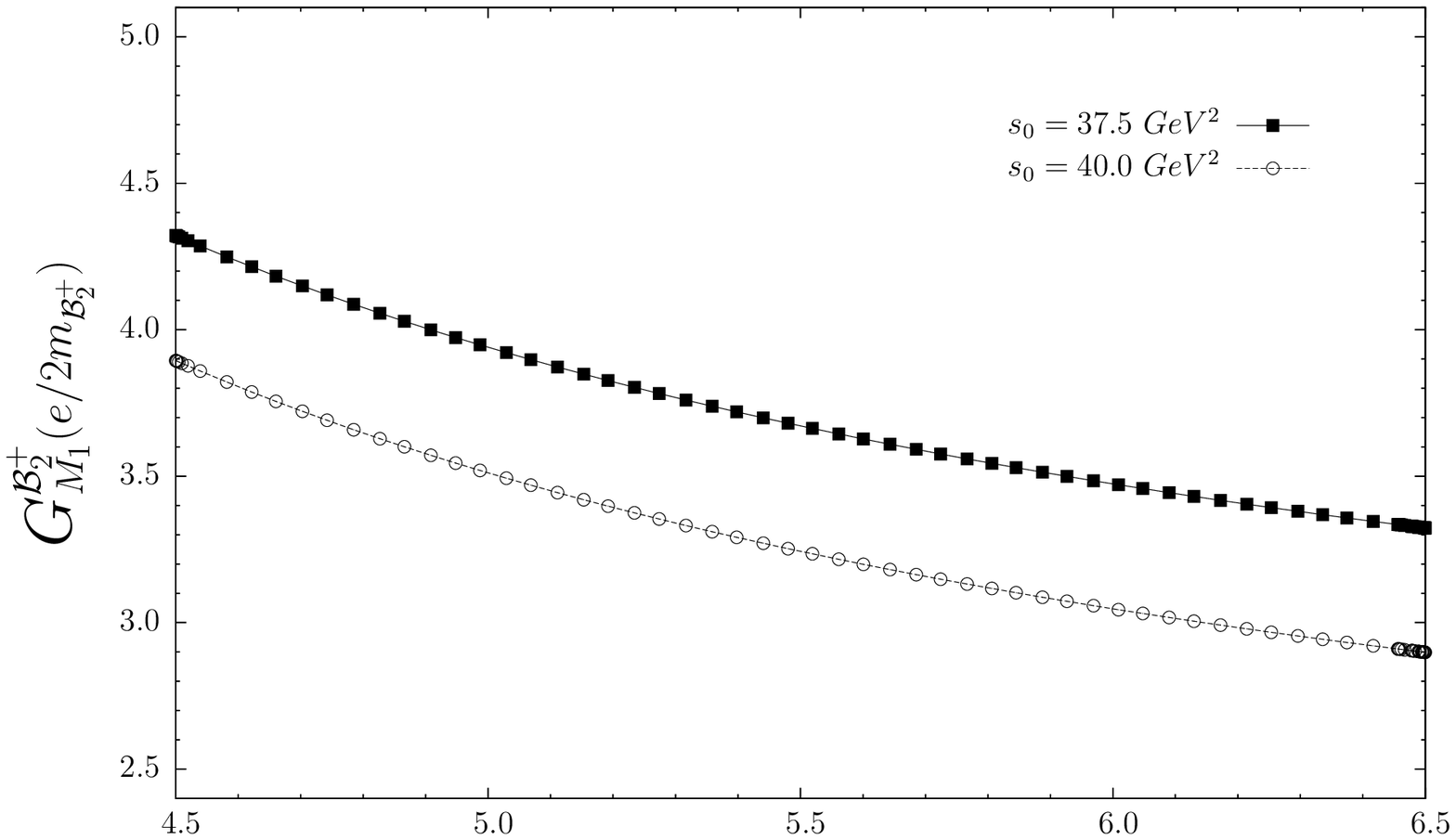}
\vskip 7.0cm
\caption{}
\end{figure}

\begin{figure} 
\vskip 3. cm
    \includegraphics{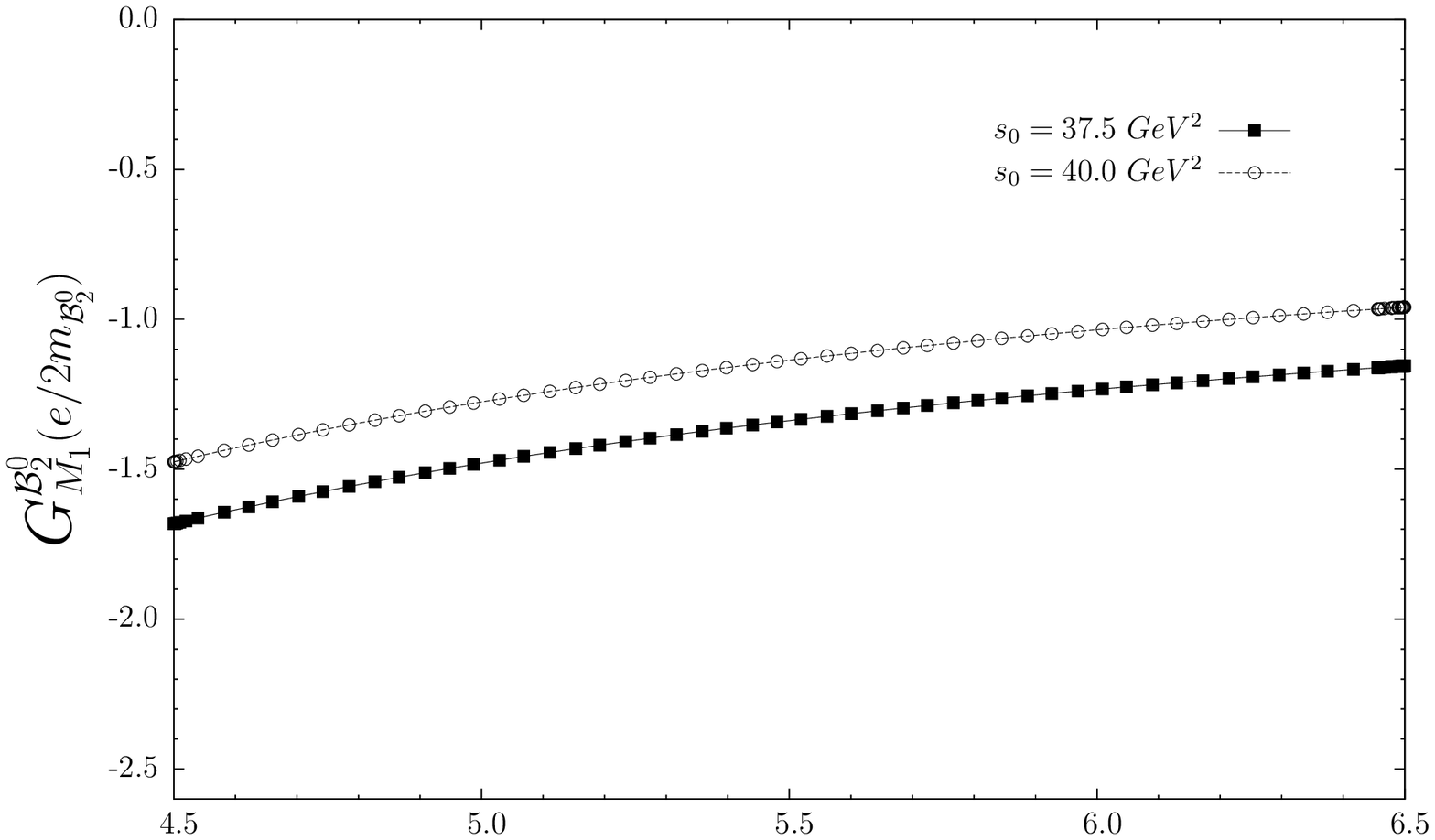}
\vskip 7.0cm
\caption{}
\end{figure}

\begin{figure}
\vskip 3. cm
    \includegraphics{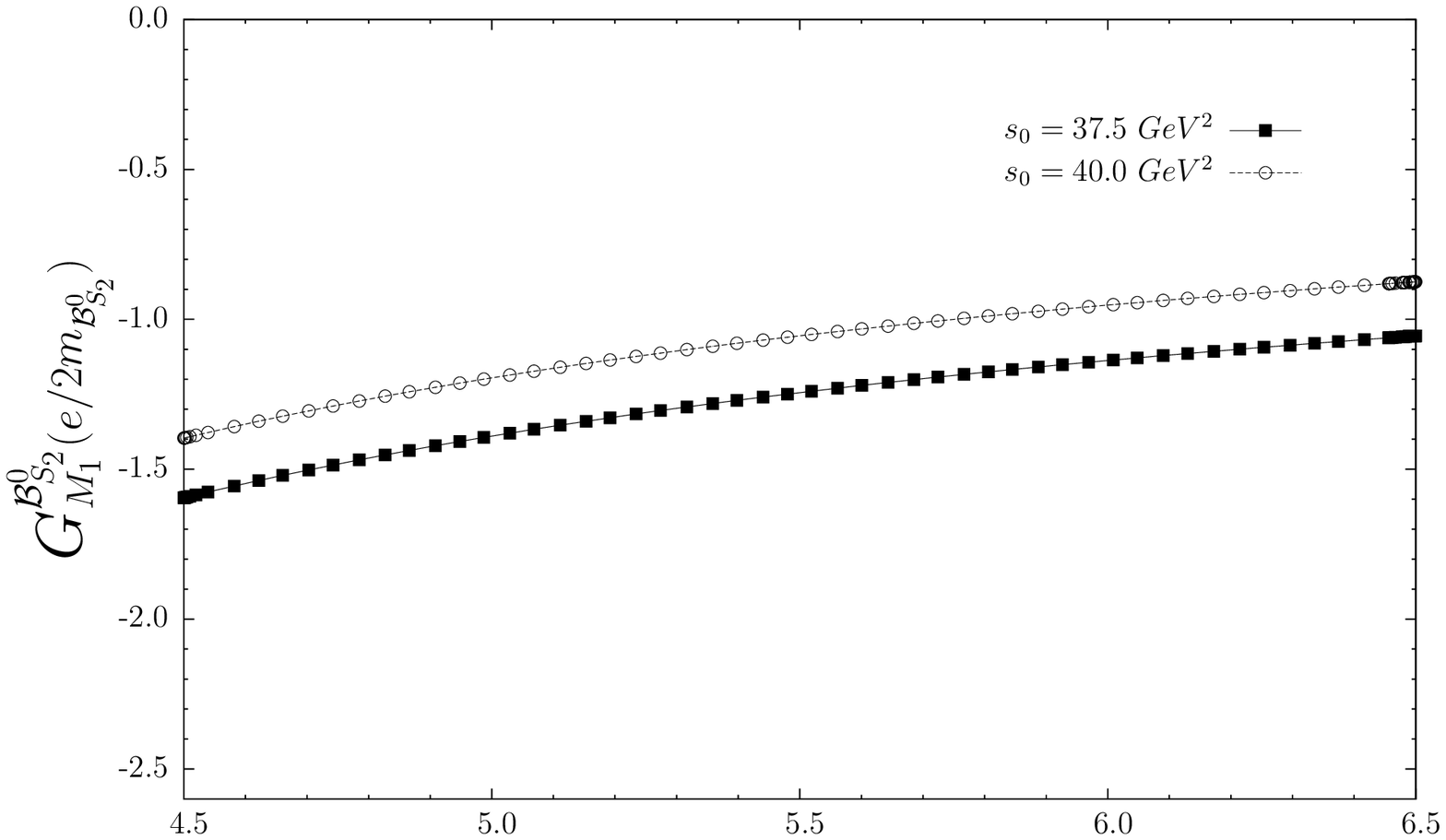}
\vskip 7.0cm
\caption{}
\end{figure}


\newpage

\begin{thebibliography}{99}

\bibitem{Rnsu01} K. Olive {\it et al.}, ParticleData Group,
  Chin. Phys. C {\bf 38}, 090001 (2014).

\bibitem{Rnsu02} E. S. Swanson,
  Phys. Rept. {\bf 429}, 243 (2006);
                S. Godfrey, and S. L. Olsen,
  Ann. Rev. Nucl. Particle Sci. {\bf 58}, 51 (2008);
                M. B. Voloshin,
  Prog. Part. Nucl. Phys. {\bf 61}, 455 (2008);
                N. Brambilla {\it et al.},
  Eur. Phys. J. C {\bf 71}, 1534 (2011);
                M. Nielsen, F. S. Navarro, S. H. Lee,
  Phys. Rept. {\bf 497}, 41 (2010).

\bibitem{Rnsu03} J. Beringen {\it et al.},
  Phys. Rev. D {\bf 86}, 010001 (2012).

\bibitem{Rnsu04} V. M. Abazov, {\it et al.},
  Phys. Rev. Lett. {\bf 99}, 172001 (2007).

\bibitem{Rnsu05} T. Aaltonen {\it et al.},
  Phys. Rev. Lett. {\bf 102}, 102003 (2009).

\bibitem{Rnsu06} T. Aaltonen {\it et al.},
  Phys. Rev. Lett. {\bf 100}, 082001 (2008).

\bibitem{Rnsu07} V. M. Abazov, {\it et al.},
  Phys. Rev. Lett. {\bf 100}, 082002 (2008).

\bibitem{Rnsu08} R. Aaij, {\it et al.},
  Phys. Rev. Lett. {\bf 110}, 151803 (2013).

\bibitem{Rnsu09} H. Sundu, K. Azizi,
  Eur. Phys. J. A {\bf 48}, 81 (2012);
                 H. Sundu, K. Azizi, Y. Sungu, and N. Yinelek,
  Phys. Rev. D {\bf 88}, 036005 (2013).

\bibitem{Rnsu10} Z. G. Wang, Z. Y. Di,
  Eur. Phys. J. A {\bf 50}, 143 (2014). 

\bibitem{Rnsu11} T. M. Aliev, M. A. Shifman,
  Phys. Lett. B {\bf 112}, 401 (1982).

\bibitem{Rnsu12} T. M. Aliev, K. Azizi, V. Bashiry,
  J. Phys. G {\bf 37}, 025001 (2010).

\bibitem{Rnsu13} T. M. Aliev, K. Azizi, M. Savc{\i},
  J. Phys. G {\bf 37}, 075008 (2010).

\bibitem{Rnsu14} V. A. Novikov, M. A. Shifman, A. I. Vainshtein, and V. I. Zakharov,
  Fortsch. Phys. {\bf 32}, 585 (1989).

\bibitem{Rnsu15} P. Ball, V. M. Braun, N. Kivel,
  Nucl. Phys. B {\bf 649}, 263 (2003). 

\bibitem{Rnsu16} C. Lorce,
  Phys. Rev. D {\bf 79}, 113011 (2009).

\bibitem{Rnsu17} A. Khodjamirian and D. Wyler,
  arXiv: 0111249 (hep--ph) (2001).

\bibitem{Rnsu18} I. I. Balitsky, V. M. Braun and A. V. Kolesnichenko,
  Nucl. Phys. B {\bf 311}, 541 (1989).

\bibitem{Rnsu19} 
  K. G. Chetyrkin, A. Khodjamirian and A. A. Pivovarov,
  Phys. Lett. B {\bf 661}, 250 (2008).

\bibitem{Rnsu20} V. M. Braun, and I. B. Filyanov,
  Z. Phys. C {\bf 48}, 239 (1990).

\bibitem{Rnsu21} V. M. Belyaev, B. L. Ioffe
  Sov. Phys. JETP {\bf 56}, 493 (1982).

\bibitem{Rnsu22} H. G. Dosch,
  Nucl. Phys. (Proc. Suppl.) B {\bf 207--208}, 312 (2010). 

\bibitem{Rnsu23} S. Narison,
  Phys. Lett. B {\bf 210}, 238 (1988).

\bibitem{Rnsu24} A. Khodjamirian, Ch. Klein, Th. Mannel, and N. Offen,
  Phys. Rev. D {\bf 80}, 114005 (2009).

\bibitem{Rnsu25} S. Narison,
  Phys. Lett. B {\bf 706},  412 (2011);
  Phys. Lett. B {\bf 707},  259 (2012);
  Phys. Lett. B {\bf 693},  559 (2010);
  Erratum ibid, {\bf 705},  544 (2011);
  Phys. Lett. B {\bf 718}, 1321 (2013);
  Phys. Lett. B {\bf 721},  269 (2013).

\bibitem{Rnsu26} J. Rohrwild,
  JHEP {\bf 09}, 073 (2007). 

\bibitem{Rnsu27} V. M. Belyaev, Ya. I. Kogan,
  Yad. Fiz. {\bf 40}, 1035 (1984).

\bibitem{Rnsu28} I. I. Balitsky, V. M. Braun, A. V. Yung,
  Yad. Fiz. {\bf 41}, 282 (1985).

\end{thebibliography}
\end {document}